\documentclass[aps,pre,preprint,superscriptaddress,showpacs]{revtex4}
\usepackage{amssymb}
\usepackage{epsfig}
\usepackage{amsmath}
\usepackage{times}

\begin{document}

\title{Synchronization in Gradient Networks}

\author{Xingang Wang}
\affiliation{Temasek Laboratories, National University of Singapore, 117508 Singapore}
\affiliation{Beijing-Hong Kong-Singapore Joint Centre for Nonlinear \& Complex Systems
(Singapore), National University of Singapore, Kent Ridge, 119260 Singapore}
\author{Ying-Cheng Lai}
\affiliation{Department of Electrical Engineering, Department of Physics and Astronomy,
Arizona State University, Tempe, Arizona 85287, USA}
\author{Choy Heng Lai}
\affiliation{Department of Physics, National University of Singapore, 117542 Singapore}
\affiliation{Beijing-Hong Kong-Singapore Joint Centre for Nonlinear \& Complex Systems
(Singapore), National University of Singapore, Kent Ridge, 119260 Singapore}

\begin{abstract}
The contradiction between the fact that many empirical networks possess
power-law degree distribution and the finding that network of heterogeneous
degree distribution is difficult to synchronize has formed a paradox in the
study of network synchronization. Surprisingly, we find that this paradox
can be well solved when proper gradients are introduced to the network
links, i.e. heterogeneous degree distribution is in favor of synchronization
in gradient networks. We analyze the general properties of gradient networks
and explore their functions in enhancing network sychronizability. Based on
these understandings, we suppose the basic principles for constructing
efficient gradient networks and propose a specific coupling scheme as
verification. Comparing to the previous asymmetric coupling schemes, the new
scheme not only possesses a much stronger synchronizability but also uses
few network information. Moreover, under the framework of gradient network,
the factors which had been employed in former studies in improving network
synchronizability can be well unified and identified. The validity of our
findings is verified by analytical estimates on the behavior of eigenvalues
as well as directed simulations on coupled nonidentical oscillators. Our
study therefore suggests that, in addition to the topology advantage,
scale-free networks also manifest their dynamical advantage given proper
gradients are considered.
\end{abstract}

\date{\today }
\author{}
\maketitle

\section{Introduction}

The study of complex networks has attracted a great deal of interest since
the discoveries of the small-world \cite{WS:1998} and scale-free \cite%
{BA:1999} properties in many natural and man-made networks \cite%
{AB:2002,NEWMAN:SIAM,BLMCH:2006}. While in initial studies the nodes and
links of a network are treated as identical, recent studies have extended to
the heterogeneous networks of scaled nodes \cite{FITNESS:2001} and weighted
links \cite{YJBT:2001} where many new features and properties are
discovered. Meanwhile, in many practical systems the scalars, which usually
reflect the different characteristics among the nodes, and the weights,
which usually characterize the information transport capacities on the
links, are closely correlated. Typical examples include the co-authorship
networks \cite{NEWMAN:2001}, where the scalar can be regarded as the number
of papers published by one researcher and the weight represents the
collaboration times between two researchers, and the world-wide airline
network (WWAN) \cite{GMTA:2005}, where the scalar can be the airport
capacity and the weight represents the amount of transportation between two
airports. In both cases the links which connect nodes of larger scalar often
assume larger weights. However, the latter is different to the former in
that in WWAN the transportation is directed and, in most cases, the mass
flows in two directions are not balanced, i.e., there exists \textit{gradient%
} on the links. The gradient networks, which are defined as \textit{the
directed graphs formed by local gradients of a scalar field distributed on
the nodes}, are ubiquitous in nature and play an important role in many
biological and technical systems \cite{TB:2004,PLZ:2005}. For example, it
has been shown that the congestion tendency of traffic networks can be
drastically reduced when gradient is considered \cite{TB:2004,PLZ:2005}.

It is well known that the collective behavior of complex systems is strongly
influenced by their underling coupling structures. One typical example is
chaos synchronization in complex networks \cite{BLMCH:2006,PRK:2001,BP:2006}%
. Compare to the regular networks, the synchronizability of both the
small-world and scale-free networks (SFN) are drastically enhanced due to
the decreased average distance \cite{WC:2002}. However, as shown by the
recent studies \cite{NMLH:2003,DTDWG:2004,ZMK:2006}, the heterogeneous
distribution of both the node degree and the link weight could suppress the
synchronizability. The contradiction between the fact that many empirical
networks have the property of heterogeneity and the finding that
synchronizability are suppressed in heterogeneous networks has formed a
paradox in the study of network synchronization. This paradox has stimulated
the searching of optimal network configuration in SFN. Specifically, for a
SFN of given topology and total coupling cost, people want to know how to
distribute the couplings could efficiently promote the synchronizability 
\cite{MZK:2005,CHAHB:2005,HCAB:2005,NM:2006}. In Ref. \cite{MZK:2005} the
authors proposed to distribute the incoming coupling strengths according to
the local information of node degree (hereafter we mark it as M-scheme). It
is found that the synchronizability is solely determined by the average
degree, independent of the degree distribution and the system size. In
particular, under the condition of uniform coupling capacity distribution,
SFN achieves its maximum synchronizability which is superior to network of
homogeneous degree distribution. In Ref. \cite{CHAHB:2005} the incoming
coupling strengths are proposed to be distributed according to the
betweenness centrality of links (hereafter we mark it as C-scheme), it is
found that synchronizability reaches its maximum only when the distributions
of these two quantities match. In both cases, the couplings are directed and
in general they are not balanced, i.e., one direction weights over another
direction.

As more and more evidences point to the important roles of gradient
couplings, an interesting question is:\ \textit{which kind of gradient will
be more efficient in improving the network synchronizability?} The answer
relies on two parallel investigations: how to set the gradient direction and
how to distribute the gradient weight. In setting the gradient direction, an
intuitive method is to let the gradient start from the larger degree node
and point to the smaller one \cite{MZK:2005,CHAHB:2005,HCAB:2005}. But
analysis of non-diagonalizable networks suggest that this setting can be
arbitrary, given no loops in the gradient network \cite{NM:2006}. In
distributing the gradient weight, the answer is even diverse and confusing,
the proposed methods range from the uniform weight distribution \cite%
{HCAB:2005} (hereafter we marked it as H-scheme) to the distributions based
on local information of node degree \cite{MZK:2005,NM:2006} and on global
information of link betweenness \cite{CHAHB:2005}. Therefore the global
picture of the function of gradient/asymmetric coupling is still not clear
and further study is needed.

In this work, we will investigate the problem of network synchronization
from the gradient network point of view and give a generic understanding to
the functions of gradient/asymmetric couplings in improving network
synchronizability. Under the framework of gradient network, not only the
previous results about asymmetric coupling can be well unified, but also the
picture of how to distribute the gradient direction and weight becomes clear
and simple. For example, now it is straightforward to figure out the
necessary conditions for heterogeneous network to have a stronger
synchronizability than homogeneous network and, more importantly, to what
extend the gradient could benefits the network synchronizability.

The rest of the paper will be organized as follows. In Section II we will
introduce the idea of gradient network and some of its basic properties.
Based on the previous experiences and the new understanding of gradient
network, In Section III we will suggest the basic criteria for constructing
optimal networks and propose a new coupling scheme as application. In
Section IV we will analysis the properties of the gradient network emerges
in the new coupling scheme and, in Section V, show its efficiency in
improving network synchronizability by the method of eigenvalue analysis. In
Section VI we will discuss the multiple effects of gradient and show how the
optimal gradient changes its value according to the network parameters. In
Section VII\ we present the simulation results on coupled nonidentical
chaotic oscillators. Finally we give the discussion and conclusion in
Section VIII.

\section{Gradient Network}

For any pair of connected nodes, there are actually two directed couplings.
When the weights in two directions are equal, we say the couplings are
symmetric, otherwise, the couplings are asymmetric. To highlight the role of
asymmetric couplings, a natural way is to separate the mutual couplings into
two parts: the symmetric part and the asymmetric part. While the symmetric
part reflects the common strength that two nodes affect each other, the
asymmetric part represents the dominant role that one node put to another
node. Extend this kind of separation to the whole network we will find that
the original asymmetrically coupled network can be separated into two
subgraphs: \textit{one undirected symmetric network and one directed
gradient network}. Both two subgraphs have the same topology as the original
network, but the coupling direction and weight have been changed, as
schematically shown in Fig. 1. Specially, the gradient network consists of
only directed links. While synchronization of symmetric network has been
well explored in previous studies, the separation of gradient network from
the asymmetric network could simplify the problem and make the analysis
easier. Now our attention solely turns to the study of gradient network and
investigate its effects on network synchronization. 
\begin{figure}[tbp]
\begin{center}
\epsfig{figure=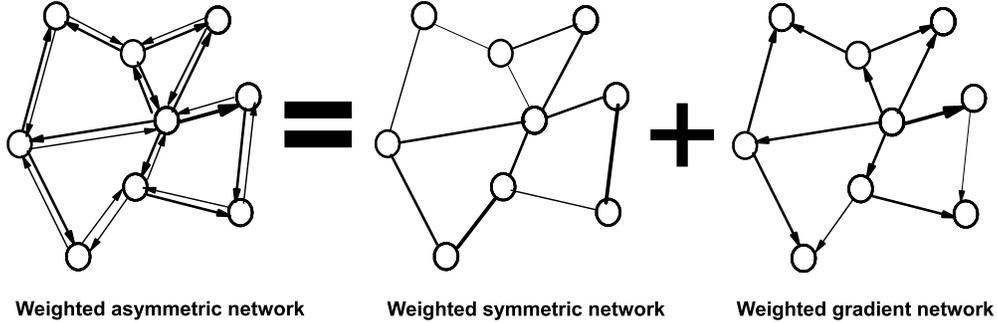,width=0.8\linewidth}
\end{center}
\caption{The schematic diagram shows how to separate an weighted asymmetric
network into a weighted symmetric network and a weighted gradient network.}
\end{figure}

Please note that the so formed gradient network is different to those
conventional ones. Conventionally gradient network is defined as the
collection of directed links pointing to each node from whichever of its
near neighbors has the highest/lowest scalar \cite{TB:2004}. For network of $%
N$ nodes and average degree $<k>$, according to the conventional definition
there will be only $N$ directed links in the gradient network. While for the
mutually coupled network, every link may have a asymmetric part and will
contributes one gradient link. Thus the gradient network generated from
coupled system actually has totally $<k>\times N/2$ directed links. By
noting that the gradients pointing to one node may have different weights
and the behavior of this node is mainly determined by the largest one (a
specific parameter will be introduced later to balance the weight
distribution), it is reasonable to reduce the gradient network from $%
<k>\times N/2$ links to $N$ links by only keeping the largest gradient
pointing to each node. In other words, for gradient network of heterogeneous
weight distribution, its main feature can be qualitatively captured by the
reduced gradient network consisting of only the important gradients. With
this simplification, the gradient network generated from coupled system will
consist with those conventional ones, therefore many results in previous
studies can be directed used here \cite{TB:2004,PLZ:2005}. Hereafter we will
only analyze the properties and functions of this kind of reduced gradient
network, while the correctness of this simplification will be verified later
by eigenvalue analysis and direct simulations.

To concrete the idea further, we describe in following the conventional way
of constructing gradient network \cite{TB:2004}. Consider complex network $%
G=G(V,E)$ which has $N$ nodes $V=\{1,....,N\}$ and $L$ links $E=\{1,...,L\}$%
. The set of edges $E$ is specified by a adjacency matrix $A=\{a_{i,j}\}$, $%
a_{i,j}=1$ if $i$ and $j$ are connected, otherwise $a_{i,j}=0$, while $%
a_{i,i}=0$. For a given node $i$ of degree $k_{i}$, the set of its neighbors
is denoted by $V_{i}=\{j\in V$ $|$ $a_{i,j}=1\}$. Consider also a scalar
field $h=\{h_{1},...,h_{N}\}$ defined on the set of nodes $V$, therefore
each node has a scalar value $h_{i}$ associated to it. For node $i$ of
degree $k_{i}$, the gradient is defined as the directed link pointing to $i$
from whichever of its $k_{i}$ neighbors has the highest scalar. If the
neighbors have the same scalar or if the node has the same scalar as all its
neighbors, the selection will be random. Gradient network is just
constructed as the collection of all these directed links. By this method,
the generated gradient network consists of $N$ nodes and $N$ links. Except
the node of the largest scalar $h_{l}=\max \{h_{1},...,h_{N}\}$, which
receives gradient from a node of smaller scalar than itself, all other nodes
only receive gradient from nodes of higher scalar than themselves.
Therefore, except the $2$-node loop formed by the largest scalar node $l$
and one of its neighbor node $j$, all the other links formed a tree
structure with nodes $l$ and $j$ at the root. It has been shown that, for
homogeneous network of random scalar distribution, the degree distribution
of gradient network follows a power law scaling, $P(k)\sim k^{-\varsigma }$,
with $\varsigma $ $\approx -1$ \cite{TB:2004}.

Comparing to those conventional gradient networks, the gradient network
generated from coupled system possesses some new features. Firstly, the node
scalar usually are not of random distribution, it has a close relation to
the properties of the node. Secondly, the gradient are weighted. Gradient
between nodes of larger scalar difference usually assumes a different weight
to that of smaller scalar difference. Finally, for sparsely connected
networks we need to consider the problem of network breaking, i.e. the
degeneration problem \cite{TB:2004}. All these new features extend the
conventional concept on gradient network and arise new problems for
investigation. Our mission is just to characterize this kind of new gradient
network and explore its functions in steering the collective behavior of
coupled networks.

\section{Constructing the Coupling Matrix}

Gradient, also known as bias, has been employed in nonlinear studies for
many years and proven to be a powerful technique in many fields such as
chaos control and chaos synchronization. For example, for coupled
oscillators on lattice it has been shown that the increase of gradient could
greatly enhance the synchronizability of the system \cite{YHX:1998}. The
asymmetric coupling schemes proposed in Refs. \cite%
{MZK:2005,CHAHB:2005,HCAB:2005} suggest that gradient can be used to complex
networks as well. Gradient has been also employed in turbulence control,
where the introduction of gradient could significantly improve the control
efficiency \cite{XHYG:1998}. While different study proposes different scheme
in setting the gradient, these schemes are nonetheless not unified. In
particular, for complex networks, people are still not clear of the proper
way of gradient configuration, i.e. how to set the gradient direction and
weight. Now, from the view point of gradient network, these studies can be
well unified and the principles for constructing optimal network can be
vividly portrayed.

By reviewing the previous studies, we can get some important clues in
constructing the matrix. (1) Based on the recent findings that 'hubs' are
firstly synchronized than 'nonhubs' and the firstly synchronized 'hubs' act
as the "core" in pattern formation of complex networks \cite{MM:2005,ZK:2006}%
, it seems plausible to build mutual links between 'hubs' in the first
place. In other words, the gradient between the hubs, if they are directly
connected, should be small. By this setting we wish to build an efficient
channel between the hubs and to form the synchronous core quickly. (2) To
guarantee that the synchronous manifold of the "core" can be efficiently
propagated to the 'nonhubs' while keeping the stability of the 'core' itself 
\cite{HCAB:2005}, it is reasonable to set the gradient start from 'hubs' and
point to 'nonhubs'. According to this requirement we can set the gradient
direction. (3) In practical systems the gradients pointing to one node
generally have different weight, neighbors of higher scalar may generate
larger gradient than those of lower scalar. To distinguish this difference,
it is reasonable to set the gradient weight being proportional to the scalar
potential. Setting in this way, the behavior of each node will more likely
to follow its larger degree neighbors. Therefore, in determining the final
synchronous state of the whole network, the high degree nodes have a larger
contribution than the smaller degree nodes. (4) For most practical systems,
each node has only limited information of the whole network, e.g. the degree
information of itself or its neighbors. Which means that, in constructing
the coupling matrix, the convenient method should employ only the local
network information \cite{MZK:2005,HCAB:2005}.

Based on the above understandings, we propose a new coupling scheme for
network synchronization. In particular, we consider networks of coupled
chaotic oscillators following equations%
\begin{equation}
\overset{.}{\mathbf{x}}_{i}=\mathbf{F}(\mathbf{x}_{i})-\varepsilon \underset{%
j=1}{\overset{N}{\sum }}G_{i,j}\mathbf{H}[\mathbf{x}_{j}],\text{ }i=1,...N,
\end{equation}%
where $\mathbf{F}(\mathbf{x}_{i})$ governs the local dynamics of uncoupled
node $i$, $\mathbf{H}[\mathbf{x}]$ is a linear coupling function and $%
\varepsilon $ is the coupling strength. $G_{i,j}$ is a zero rowsum coupling
matrix with off diagonal entries read%
\begin{equation}
G_{i,j}=\frac{A_{i,j}k_{j}^{\beta }}{\sum_{j=1}^{N}L_{i,j}k_{j}^{\beta }},%
\text{ }i,j=1,...N,  \label{matrix}
\end{equation}%
with $A$ the adjacency matrix defined in Sec. II, $k_{j}$ denotes the degree
of node $j$, and $\beta $ is a tunable parameter which will be used to
adjust the gradient weight distribution. The diagonal entries are unit $%
G_{i,i}=1$.

\section{The Properties of Gradient Network}

We first investigate the properties of the gradient network formed by Eq. %
\ref{matrix}. For a pair of connected nodes $i$ and $j$, denote $k_{\iota }$
($k_{j}$) the degree of node $i$ ($j$) and $V_{i}$ ($V_{j}$) the set of its
neighbors, then the gradient from $i$ to $j$ reads 
\begin{equation}
\Delta G_{j,i}=G_{j,i}-G_{i,j}=\frac{k_{i}^{\beta }}{\sum_{l\in
V_{j}}k_{l}^{\beta }}-\frac{k_{j}^{\beta }}{\sum_{l^{\prime }\in
V_{i}}k_{l^{\prime }}^{\beta }}=\frac{1}{C_{i,j}}[k_{i}^{\beta
}\sum_{l^{\prime }\in V_{i}}k_{l^{\prime }}^{\beta }-k_{j}^{\beta
}\sum_{l\in V_{j}}k_{l}^{\beta }],  \label{gradient}
\end{equation}%
with $C_{i,j}=\sum_{l\in V_{j}}\sum_{l^{\prime }\in V_{i}}k_{l}^{\beta
}k_{l^{\prime }}^{\beta }$. With the language of gradient network, it is
straightforward to define the node scalar as follows 
\begin{equation}
h_{i}=k_{i}^{\beta }\sum_{l\in V_{i}}k_{l}^{\beta }.  \label{scalar}
\end{equation}%
This scalar will be used to determine the gradient direction, i.e. from
higher scalar to lower scalar; while the gradient weight is determined by
the scalar potential $\Delta h_{i,j}=h_{i}-h_{j}$ and $C_{i,j}$, which
depend on the degree situation of the node and its the neighbors. For SFN
generated via the BA model \cite{BA:1999}, there is no degree correlation
between the nodes \cite{NEWMAN:2002}, i.e. the chance to find a large degree
neighbor or a small degree neighbor is the same for any given node.
Specifically, we have for this kind of network the relation $\sum_{l\in
V_{i}}k_{l}^{\beta }\sim k_{i}$. Substitute this relation into Eq. \ref%
{scalar} we have 
\begin{equation}
h_{i}\sim k_{i}^{1+\beta }.  \label{scalar2}
\end{equation}%
It can be found that, for positive value of $\beta $, larger degree node
generally possesses a larger value of $h_{i}$ and the gradient is started
from the larger degree node and point to the smaller degree node, while the
contrary situation occurs when $\beta <0$. (Note that with a slim chance the
gradient may flow from smaller degree node to larger degree node. This
situation happens when a smaller degree nodes has very large degree
neighbors. However, this chance is very small and, statistically, the
relation Eq. \ref{scalar2} is still valid, specially for network of dense
connections.) Therefore, we can control the gradient direction by changing
the sign of $\beta $, while control its weight, which is proportional to the
scalar difference as shown in Eq. \ref{gradient}, by its absolute value $%
\left\vert \beta \right\vert $.

\begin{figure}[tbp]
\begin{center}
\epsfig{figure=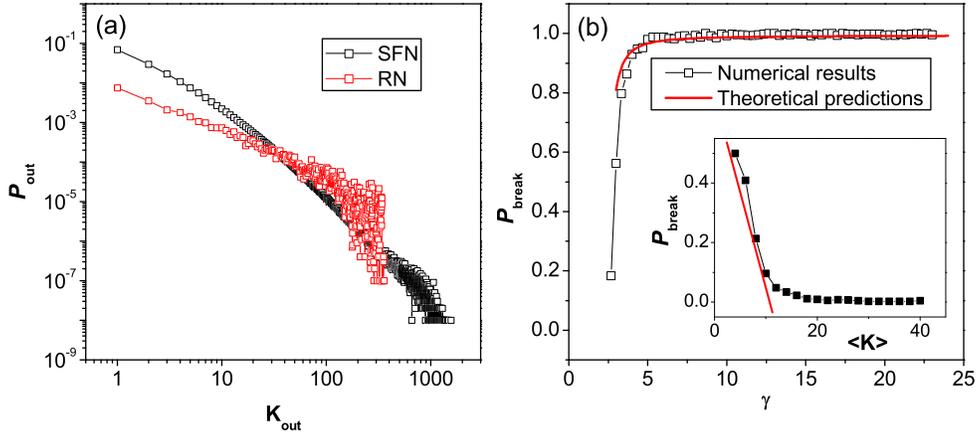,width=0.8\linewidth}
\end{center}
\caption{(Color online). For network of $N=2^{10}$ nodes and mean degree $%
<k>=6$. (a) The out-degree distribution of the gradient networks generated
on SFN and RN substrates. Both distributions follow power-law scaling with $%
\protect\varsigma \approx -2.8$ for SFN substrate and $\protect\varsigma %
\approx -1$ for RN substrate. (b) The probability of network breaking as a
function of degree heterogeneity $\protect\gamma $. Inset plots the breaking
probability as a function of the mean degree $<k>$. It is found that network
tends to be broken as $\protect\gamma $ increases or $<k>$ decreases. Each
data is an average result over $100$ realizations.}
\label{fig:gradient}
\end{figure}

The constructed gradient network consists of $N$ nodes and $<k>\times N/2$
directed links. To simplify the analysis, we assume $\beta \rightarrow
\infty $ and discuss the reduced gradient network which consists of only $N$
links. As we discussed in Sec. II, this reduction will not affect the main
results we get for gradient network, but will make the analysis much easier.
Now we argue that the formed gradient network is of forest structure (for
obvious reason, we ignore the $2$-node loop formed by the largest degree
node and one of its neighbors) and its out-degree follows a power law degree
distribution. To prove the forest structure, assume that on the contrary,
there is a closed path $\Phi =\{\Delta G_{2,1},\Delta G_{3,2},...,\Delta
G_{l,l-1}\},l\geq 3$ made up only of directed edges selected from the
gradient network. Let $j$ be the node on this path for which $h_{j}$ $%
=min\{h_{1},h_{2},...,h_{m}\}$. Node $j$ has exactly two neighbors on $\Phi $%
, nodes $j\pm 1$, but only one gradient, $\Delta G_{j,j+1}$ or $\Delta
G_{j,j-1}$, pointing to $j$. Since $h_{j}<h_{j\pm 1}$, both of the neighbors 
$j\pm 1$ will have their gradient edges pointing into $j$. Since there are
two edges, $(j-1,j)$ and $(j,j+1)$, and only one gradient edge will be
accepted by $j$, one of the edges must not be a gradient edge, and thus the
loop is not closed, in contradiction with the assumption that there is a
loop with only gradient edges.

We go on to estimate the out-degree distribution. For node $i$ which has
degree $k_{i}$, assume node $l\in V_{i}$ is one of its neighbors, then the
probability for $l$ to receive gradient from $i$ is determined by two
elements: the probability of $h_{i}>h_{l}$ and the probability$\ $that $h_{i}
$ has the largest scalar among the neighbors of $l$, i.e. $h_{i}=\allowbreak
\max \{h_{j},$ $j\in V_{l}\}$. Due to the zero degree correlation, the
chance to find a connection between $i$ and $l$ is $p_{1}=k_{i}/(N-1)$;
according to Eq. \ref{scalar2}, the probability for $h_{i}>h_{l}$ equals
that of $k_{i}>k_{l}$ (here we discuss only the case of $\beta >0$, while
the analysis for $\beta <0$ is the similar), which is $p_{2}=\int_{k_{\min
}}^{k_{i}}P(k)dk$; the probability that $h_{i}$ is the largest scalar among
the neighbors of $l$ is $p_{3}==1-(k_{l}-1)p_{3}^{\prime
}=1-(k_{l}-1)\int_{k_{i}}^{k_{\max }}P(k)dk$, where $p_{3}^{\prime }$
represents the probability that one link of $l$ connects to some node of
larger degree to $i$. Therefore, the probability for $l$ to contribute an
out-degree to $i$ is%
\begin{equation}
p_{l}=p_{1}p_{2}p_{3}=k_{i}/(N-1)\cdot \int_{k_{\min }}^{k_{i}}P(k)dk\cdot 
\left[ 1-(k_{l}-1)\int_{k_{i}}^{k_{\max }}P(k)dk\right] .  \label{p1}
\end{equation}%
According to the network growth algorithm \cite{BA:1999}, we have $k_{\max
}\approx m\times N^{\frac{1}{\gamma -1}}$ and $k_{\min }=m$ ($m$ is the
number of links associated to the new node in the BA growth model). Noting
that the contribution of out-degree may come from any node of degree smaller
than $k_{i}$, substituting the relation $P(k)=Ck^{-\gamma }=(\gamma
-1)m^{\gamma -1}k^{-\gamma }$ into Eq. \ref{p1}, we get the probability for
each link of $i$ to become an out-degree link is $\int_{k_{\min
}}^{k_{i}}p_{l}NP(l)dl$. Finally, the totoal number of out-degree for node $i
$ is 
\begin{eqnarray}
k_{out} &=&k_{i}\cdot \int_{k_{\min }}^{k_{i}}p_{l}NP(l)dl\approx
\int_{m}^{k_{i}}\left\{ k_{i}\cdot \int_{m}^{k_{i}}P(k)dk\cdot \left[
1-(l-1)\int_{k_{i}}^{k_{\max }}P(k)dk\right] \right\} Cl^{-\gamma }dl  \notag
\\
&=&C^{2}k_{i}\cdot \int_{m}^{k_{i}}\left\{ k_{i}\cdot
\int_{m}^{k_{i}}k^{-\gamma }dk\cdot \left[ 1-C(l-1)\int_{k_{i}}^{k_{\max
}}k^{-\gamma }dk\right] \right\} l^{-\gamma }dl
\end{eqnarray}%
Denoting $\int_{m}^{k_{i}}k^{-\gamma }dk=a$ and $\int_{k_{i}}^{k_{\max
}}k^{-\gamma }dk=b$, we have%
\begin{eqnarray}
k_{out} &=&C^{2}k_{i}\cdot \int_{m}^{k_{i}}\left\{ k_{i}\cdot a\cdot \left[
1-C(l-1)b\right] \right\} l^{-\gamma }dl  \notag \\
&=&aC^{2}k_{i}\cdot \left\{ (1+Cb)a-Cb\frac{m^{2-\gamma }-k_{i}^{2-\gamma }}{%
\gamma -2}\right\} 
\end{eqnarray}%
For $k_{i}\lesssim k_{\max }$, $Ca\approx 1$ and $Cb\approx 0$, we have%
\begin{equation}
k_{out}\approx k_{i}  \label{kout}
\end{equation}%
Therefore the out-degree distribution of the gradient network is the same to
the degree distribution of SFN, which is 
\begin{equation}
p_{out}\sim P(k)\sim k_{i}^{-\gamma }.  \label{outdegree}
\end{equation}%
The exponent we get from numerical simulation matches this prediction very
well, as shown in Fig. 2(a) where the fitted exponent is $\varsigma \approx
-2.8$. Following the similar deduction, we can also proof that the
out-degree distribution of random network also follows a power law
distribution, but with exponent $\varsigma =-1$. This relation is verified
by numerical simulation again and the result is plotted in Fig. 2(a) by
another curve.

Using a similar reasoning as for the forest structure, we can prove that
there is no continuous path which connects two local maxima of the scalar
field $h$. This means that for each tree of the gradient network there is
only one local maximum scalar, and it is the only node which forms a $2$%
-node loop with one of its neighbors. As a consequence, the number of trees
in the forest equals the number of local maxima of the scalar field $h$. In
other world, if there are more than one tree appear in the forest, the
gradient network will be broken into disconnected small graphs. This kind of
breaking phenomenon is crucial to network synchronization, since once the
network is disconnected, it will never be synchronized whatever how large
the coupling strength is. Therefore, to fully explore the function of
gradient, we also need to estimate the breaking risk it may induce.
Specifically, we want to know which kind of network could bear a larger
gradient. In following we argue that \textit{densely connected heterogeneous
networks are more capable of larger gradient than sparsely connected
homogeneous networks}. For SFN of degree heterogeneity $\gamma $, the
largest node has about $k_{\max }\approx \frac{<k>}{2}N^{\frac{1}{\gamma -1}%
} $ links. Therefore the chance for the two leading nodes to be directly
linked is about $P_{con}\approx 2k_{\max }/N$. In other words, the chance
for them to be indirectly linked will be $P_{break}=1-P_{con}$. As the
forest will broken into at least two trees when the two leading nodes are
not directly connected, we have the network breaking probability 
\begin{equation}
P_{break}=1-2k_{\max }/N\approx 1-<k>N^{\frac{2-\gamma }{\gamma -1}}.
\label{break}
\end{equation}%
This is only an estimation of the breaking probability, since the indirect
connection of other relatively larger degree nodes, instead of the two
largest ones, may also break down the forest in some cases. Such chance,
however, is very small comparing to $P_{break}$ and can be neglected,
especially for networks\ of larger $\gamma $. (To construct networks of
variable $\gamma $, we adopt the model proposed in Ref. \cite{DM:2002}, i.e.
set a tunable parameter $B$ to each node and adjust the preferential
attachment function to be $p\sim (k_{i}+B)/\sum_{j}(k_{j}+B)$. The scaling
exponent $\gamma $ is then given by $\gamma =3+B/m$, with $m$ the number of
new links associated to each new node in network growth.) Numerical results
on the breaking probability as functions of $\gamma $ and $<k>$ are plotted
in Fig. 2(b). It can be found that the predictions fit the numerical results
reasonably well, especially in regions of larger $\gamma $ and smaller $<k>$.

\section{Synchronizability Analysis}

We now analyze the function of gradient network to network synchronization.
The synchronizability of coupled networks can be evaluated by the method of
master stability function (MSF) \cite{PC:1998,BP:2002}, if the eigenvalues
are reals, or by the method of eigenvalue analysis \cite{YHX:1998}, if the
eigenvalues are complex values. These methods tell us that the problem of
synchronizability can be divided into two separating issues: the stability
of the single dynamics $\mathbf{F}(\mathbf{x})$ and the distribution of
eigenvalues of the coupling matrix $G$. For most systems, the single
dynamics is stable within a certain range in the parameter space, $\sigma
\in \lbrack \sigma _{1},\sigma _{2}]$. The network is synchronizable iff all
the eigenvalues except the one $\lambda _{1}=0$, which corresponding to the
synchronous manifold, can be contained within this range after a linear
scaling, i.e. $\lambda _{N}/\lambda _{2}\leqslant \varepsilon \sigma
_{2}/\sigma _{1}$, with $\lambda _{N}$ the largest and $\lambda _{2}$ the
smallest positive eigenvalues, respectively. In other words, the quantity of
synchronizability can be described by the eigenratio $R=\lambda _{N}/\lambda
_{2}$, with a smaller $R$ represents a stronger synchronizability.
Meanwhile, when network is synchronized, larger $\lambda _{2}$ usually means
smaller coupling cost since $\varepsilon >$ $\sigma _{1}/\lambda _{2}$.

The eigenvalues of asymmetric matrix $G$ are usually complex values \cite%
{YHX:1998,HCAB:2005}, but for the coupling matrix $G$ constructed in Eq. \ref%
{matrix}, they are reals. Noticing that the coupling matrix can be written
as $G=QLD^{\beta }$, with $D=diag\{k_{1},k_{2},...k_{N}\}$ the diagonal
matrix of degrees and $Q=diag\{1/\sum_{j}L_{1,j}k_{j}^{\beta
},...1/\sum_{j}L_{N,j}k_{j}^{\beta }\}$ the normalization factors for rows
of $G$. From the following identity 
\begin{equation}
\det (QLD^{\beta }-\lambda I)=\det (Q^{1/2}D^{\beta /2}LD^{\beta
/2}Q^{1/2}-\lambda I)
\end{equation}%
we can find that the eigenvalues of the asymmetric matrix $G$ are the same
as that obtained from the symmetric matrix $H=$ $Q^{1/2}D^{\beta
/2}LD^{\beta /2}Q^{1/2}$, which are real and nonnegative values.

From the viewpoint of gradient network, the previous coupling schemes \cite%
{MZK:2005,CHAHB:2005,HCAB:2005} can be well unified and the role of gradient
can be easily identified. In M-scheme the gradient is generated according to
the degree difference and adjusted via parameter $\beta _{M}$, i.e. $%
h_{i}=1/k_{i}^{\beta }$ and $\Delta G_{j,i}=(k_{i}^{\beta }-k_{j}^{\beta
})/(k_{i}^{\beta }k_{j}^{\beta })$. A negative value of $\beta _{M}$
represents that the gradient flows from the smaller degree node to the
larger degree node, while for positive $\beta _{M}$ the gradient flows in
the opposite direction. As reported In Ref. \cite{MZK:2005} and repeated in
Fig. 3(a), the maximum synchronizability happens at $\beta _{M}\approx 1$.
Since the gradient increases its weight as $\beta _{M}$ increases from $0$,
it is of certain surprise to find that, instead of enhancement, larger
gradient will suppress synchronizability when $\beta _{M}>1$. Another
intriguing observation is the sharp change of eigenratio $R$ as $\beta $
varies: $R\approx 2\times 10^{3}$ at $\beta _{M}=-1$ while $R\approx 6$ at $%
\beta _{M}=1$. While the optimization at $\beta _{M}\approx 1$ can be
understood by the heterogeneous distribution of the coupling capacity among
the nodes (at $\beta _{M}=1$ the coupling capacity is the same for all
nodes) and the decreased total coupling cost as $\beta _{M}$ increases from
the optimal value (as $\beta _{M}$ increases from $1$ the total coupling of
the network will be decreased), the sharp change of $R$ demonstrates the
nontrivial effect of gradient played in network synchronization. In Fig.
3(a) we also plotted the variations of the eigenratio as a function of the
gradient for C-scheme, where gradient is generated by the betweenness
difference between the connected nodes, and for H-scheme, where gradient is
generated by the aging difference. For C-scheme, the gradient weight is
adjusted by parameter $\beta _{C}$, with negative values represent that the
gradient flows from node of smaller betweenness to node of larger
betweenness, while the opposite happens for positive $\beta _{C}$. Again,
the absolute value of $\beta _{C}$ represents the weight of gradient. The
behavior of $R$ is quite similar to that of M-scheme, i.e. optimal
configuration exists at around $\beta _{C}\approx 1$ while larger gradients
suppress synchronizability, except that the variation of eigenratio is much
slow than that of M-scheme. For H-scheme the gradient\ weight is adjusted
via the parameter $\beta _{H}$, with negative values represent that the
gradient flows from the 'older' (larger degree) node to the 'younger'
(smaller degree) node, while opposite happens for positive $\beta _{H}$.
Again, the absolute value of $\beta _{H}$ represents the weight of gradient.
Different to the former two schemes, it is found in Fig. 3(a) that for
H-scheme the eigenratio $R$ monotonically decreases as $\beta _{H}$
decreases from $1$ to $-1$, or, similarly, as the gradient from 'older' to
'younger' increases its weight. Noticing that $\beta _{H}=0$ equals the
situation $\beta _{M}=1$ in M-scheme, it seems that for H-scheme the
increase of gradient will always enhance synchronizability. However, from
the view point of gradient network, the increase of gradient in H-scheme may
induce the breaking problem which will suppress synchronization. Meanwhile,
a 'younger' node receives gradients \textit{equally} from all its 'older'
neighbors despite their detail difference, this may confuse the target
synchronous state to which the 'younger' node should follow \cite{NMLH:2003}%
. Therefore the gradient functions are not fully explored in H-scheme.

As a comparison, we also plot in Fig. 3(a)\ the result of the new scheme
described by Eq. \ref{matrix}. For $\beta <0$, the gradient flows from
smaller degree node to larger degree node and the opposite happens when $%
\beta >0$. The weight of the gradient is adjusted by the absolute value of $%
\beta $. It can be found that, as $\beta $ increases, the eigenratio $R$ 
\textit{monotonically} decreases from large values to small values. As we
will show later, the smallest value that $R$ can reach is only determined by
the largest eigenvalue $\lambda _{N}$, which is almost constant for
different coupling schemes. (In the extreme situation of $\beta \rightarrow
\infty $, the coupling matrix takes the form of the reduced gradient
network. There will be only two eigenvalues, $\lambda _{N}$ and $\lambda
_{2} $, while $\lambda _{2}$ equals $1$.) When $\beta =0$, we recover to the
situations of $\beta _{H}=\beta _{C}=0$ and $\beta _{M}=1$ used in the
previous schemes, respectively. Significantly, the maximum synchronizability
at $\beta _{H}=-1$ can be achieved at around $\beta \approx 5$, while $R$
can be further decreased as $\beta $ increases in the new scheme. In this
sense, we say that the new scheme is more efficient in employing gradient
than the other schemes. 
\begin{figure}[tbp]
\begin{center}
\epsfig{figure=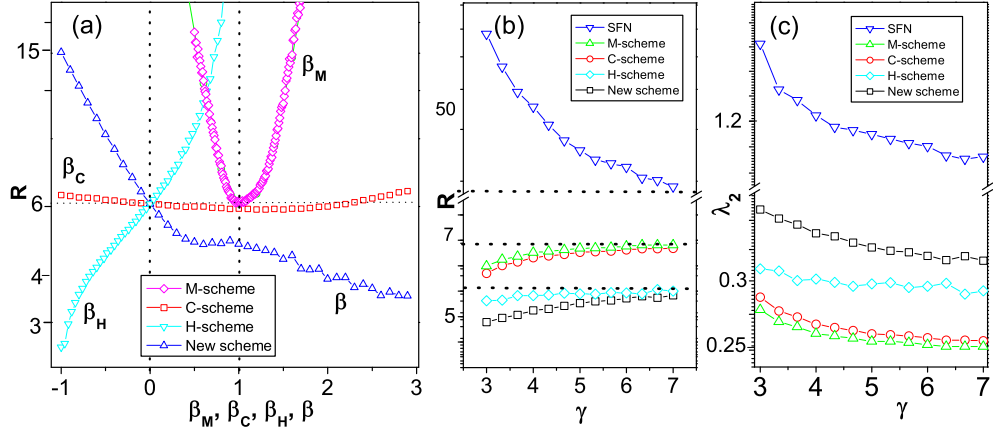,width=0.8\linewidth}
\end{center}
\caption{(Color online). For the same SFNs as in Fig. 2. (a) The variations
of eigenratio $R$ as functions of the gradient parameters, $\protect\beta %
_{M}$, $\protect\beta _{C}$, $\protect\beta _{H}$, and $\protect\beta $, for
different coupling schemes. (b) The eigenratio $R$ as a function of degree
heterogeneity $\protect\gamma $ for different coupling schemes. The three
dashed lines represent the eigenratio of RN under the situation of, from top
to bottom, without gradient, with M-scheme, and with the new scheme of Eq. 
\protect\ref{matrix}, respectively. (c) The variation of the $\protect%
\lambda _{2}$ as a function of heterogeneity for different schemes. Each
data is averaged over $50$ realizations.}
\end{figure}

Noticing the fact that increasing degree heterogeneity could suppress
synchronizability \cite{NMLH:2003}, we go on to compare the
synchronizabilities of the different schemes as a function of heterogeneity.
The variation of $R$ as a function of $\gamma $ is plotted in Fig. 2(b)
together with four reference configurations: SFN and RN without gradient, RN
of M-scheme, and RN of the new coupling scheme. To make the comparison fair,
we adopt $\beta _{M}=1$ and $\beta _{C}=1$, where the maximum
synchronizabilities are reached for the corresponding schemes. To make a
fair comparison between the new scheme and the H-scheme, we adopt $\beta
_{H}=-0.5$ and $\beta =1.5$, since under this setting the total gradient is
equal for these two schemes. It can be found that the new scheme has a clear
advantage over all the other schemes. It is also found that, under the new
scheme, networks of higher heterogeneity shows a much clear advantage over
the homogeneous ones, while for the other schemes the advantage is
relatively weak. Based on this observation, we say that the new scheme not
only efficiently enhances the synchronizability of SFNs, as compared to
their original configurations, but also makes SFNs prominently superior to
RN. This finding provides a more stronger explanation to the paradox of
network synchronization \cite{MZK:2005}. Another quantity used to
characterize synchronizability is $\lambda _{2}$, the second smallest
eigenvalue. Since a larger value of $\lambda _{2}$ usually represents a
smaller coupling cost in reaching the global synchronization, i.e. $%
\varepsilon >$ $\sigma _{1}/\lambda _{2}$. The variation of $\lambda _{2}$
as a function of heterogeneity is plotted in Fig. 3(c), it is found that the
new scheme has a larger value of $\lambda _{2}$ in comparison with the other
schemes.

\section{Understanding the Multiple Effects of Gradient}

To explore the underlying mechanisms behind the new scheme, and also to
manifest the basic principles we proposed for network construction in Sec.
III, we go on to characterize the gradient network by other two quantities:
the distribution of gradient weight and the distribution of eigenvalues. The
gradient weight, as described by Eq. \ref{gradient}, is some value between $%
\Delta G_{\min }\approx 0$, where the connected nodes have the similar
scalar, and $\Delta G_{\max }\approx 1$, where the link connect the largest
degree and the smallest nodes. In Fig. 4(a) we plot the weight distributions
of the gradient networks for all the coupling schemes. A clear difference is
that the weight distribution of the new scheme has a long tail. According to
Eq. \ref{matrix}, a larger $\Delta G_{i,j}$ represents a larger degree
difference between the connected nodes. This is in accordance with the
second principle in network construction, i.e. the gradient weight should be
proportional to the the scalar potential. As a comparison, in other schemes
the node receives gradients from it neighbors in a relatively mean fashion.
For example, In H-scheme, the 'younger' node receives equal gradient from
all the 'older' neighbors, while disregards the degree difference among
them. The advantage of the new scheme is also reflected in the
eigenspectrum. In Fig. 4(b)\ we plot the eigenvalue distributions for
different coupling schemes. It can be found that, while the largest
eigenvalue $\lambda _{N}$ is similar for all the schemes, the new scheme is
distinct with a larger value of $\lambda _{2}$ and an absolutely higher
probability around $\lambda =1$. This property makes the eigenvalues to be
restricted within a tight region around $\lambda =1$, and make the value of
eigenratio $R$ be quickly decreased as $\beta $ increases. The extreme
situation will be all the eigenvalues, except $\lambda _{N}$ and $\lambda
_{1}$, equal $1$, where the maximum synchronizability is achieved and $%
R=\lambda _{N}$ .

From Figs. 4(a)\ and (b), it seems that increasing gradient will
monotonically increases the synchronizability. However, as shown in Fig.
2(b), too large gradients may suppress synchronizability, since gradient
could induce the risk of network breaking. Now we show this phenomenon via
eigenvalue analysis. In Fig. 4(c) we plot eigenratio $R$ versus gradient $%
\beta $ for networks of different heterogeneities. It is found that, as the
heterogeneity decreases (i.e. increase the value of $\gamma $), in the
region of $\beta $, the gradient gradually changes its role from enhancing
synchronizability to suppressing synchronizability. That is, for networks of
lower heterogeneity and smaller mean degree, there exist an optimal gradient
at $\beta _{o}$. When $\beta <\beta _{o}$, increasing gradient will enhance
synchronizability, but when $\beta <\beta _{o}$, the opposite happens. The
position where this transition appears depends on the values of $\gamma $
and $<k>$, for densely connected SFN the value $\beta _{o}$ will be
significantly delayed. The breaking effect can be further understood from
Fig. 4(d), where we plot the eigenratio $R$ for a number of network
realizations for fixed $\gamma $ and $<k>$. It is found that, for smaller
gradient $\beta <\beta _{o}$, $R$ keeps on small values, representing that
the network is still well connected. In this region, increase $\beta $ will
enhance the synchronizability; but for larger gradient $\beta >\beta _{o}$, $%
R$ intermittently jumps to very large values, reflecting that the network is
broken in some realizations. Therefore in this region increase $\beta $ will
suppress the synchronizability.

\begin{figure}[tbp]
\begin{center}
\epsfig{figure=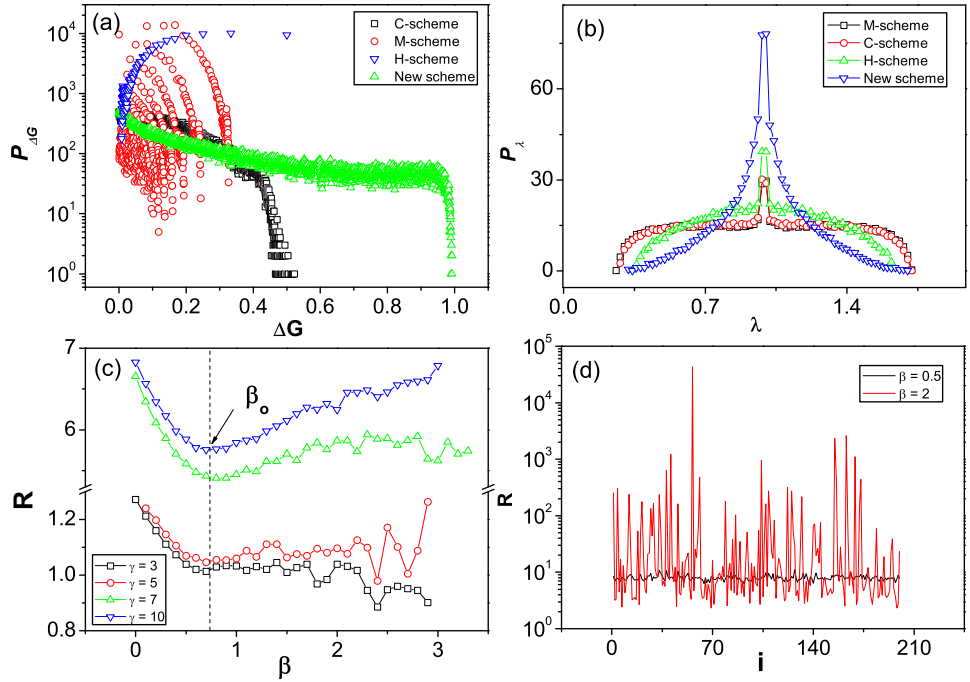,width=0.8\linewidth}
\end{center}
\caption{(Color online). For the same SFN as in Fig. 2. Comparing (a) the
distributions of gradient weight $\Delta G$ and (b) distributions of
eigenvalues among the different coupling schemes. (c) For the new scheme,
the variation of $R$ as a function of $\protect\beta $ for networks of
different heterogeneities. Optimal gradient happens around $\protect\beta %
_{o}\approx 0.7$ when $\protect\gamma =10$. (d) The distribution of $R$ as a
function of the network realizations for different $\protect\beta $ at $%
\protect\gamma =10$. }
\label{fig:eigen}
\end{figure}

\section{Numerical Simulations}

We now provide the results of direct simulations. It was shown that,
although the MSF\ method was proposed for complete synchronization of
coupled identical systems, the eigenratio $R$ could still provide a
qualitative description for the collective behaviors, e.g. phase
synchronization, of coupled nonidentical systems \cite%
{CHAHB:2005,MZK:2005,ZK:2006}. We employ SFN of nonidentical chaotic R$%
\overset{..}{\text{o}}$ssler oscillators as the model. The dynamics of a
singular oscillator reads $\mathbf{F}_{i}(\mathbf{x}_{i})=[-\omega
_{i}y_{i}-z_{i},\omega _{i}x_{i}+0.15y_{i},z_{i}(x_{i}-8.5)+0.4]$, with $%
\omega _{i}$ the natural frequency of oscillator $i$, which is randomly
assigned in range $[0.9,1.1]$. The coupling form is $\mathbf{H}(\mathbf{x})=%
\mathbf{x}$. The degree of synchronization in this model can be
characterized by monitoring the amplitude $A$ of the mean field $%
X=\sum_{i=1}^{N}x_{i}/N$ \cite{MZK:2005,ZK:2006}. For small coupling
strength, $X$ oscillates irregularly and $A$ is approximately zero,
reflecting a smaller degree of synchronization; while $X$ oscillates
regularly and $A$ increases sharply as coupling strength exceeds a critical
value, reflecting a larger degree of synchronization. In Fig. 5(a) we plot
the behavior of $A$ as a function of the coupling strength $\varepsilon $
for all the coupling schemes, it is found that the new scheme has a clear
advantage over the other schemes even for smaller $\beta $, especially in
the small the region of small couplings. To demonstrate the positive effect
that gradient plays in the new scheme, we plot in Fig. 5(b)\ the behavior of 
$A$ as a function of the gradient degree $\beta $. Again, as predicted by
the eigenvalue analysis, $A$ increases monotonically as $\beta $ increases.

\begin{figure}[tbp]
\begin{center}
\epsfig{figure=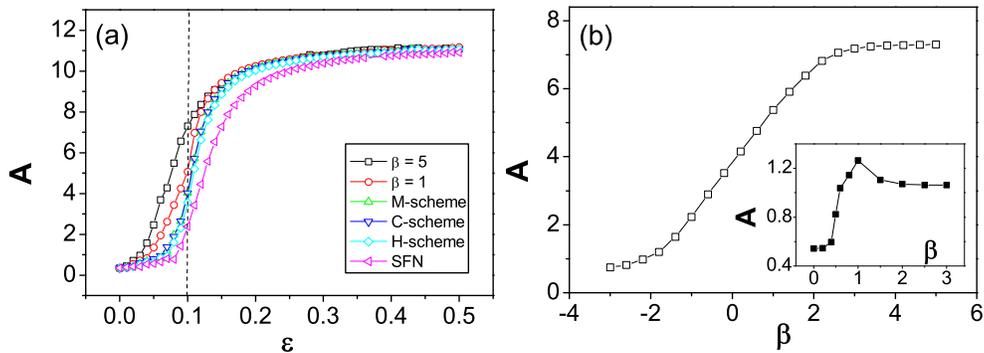,width=0.8\linewidth}
\end{center}
\caption{(Color online). For SFNs of $N=2^{10}$ nodes and mean degree $%
<k>=10 $, directed simulation of coupled nonidentical R$\protect\overset{..}{%
o}$ssler oscillators. (a)\ The mean field amplitudes as a function of
coupling strength for different coupling schemes. The parameters are $%
\protect\beta _{M}=1$ for M-scheme, $\protect\beta _{C}=1$ for C-scheme, $%
\protect\beta _{H}=-0.5$ for H-scheme, $\protect\beta =1.5$ and $5$ for the
new scheme. The total gradient in H-scheme when $\protect\beta _{H}=-0.5$
and $-1$ equal that of the new scheme when $\protect\beta =1.5$ and $5$,
respectively. (b) For the new scheme, fixing $\protect\varepsilon =0.1$, the
variation of mean field amplitude as a function of gradient. Each data is
averaged over $10$ network realizations. Inset is plotted for sparsely
linked homogeneous network where each data is averaged over $1000$ network
realizations.}
\label{fig:simulations}
\end{figure}

To show the breaking effect, we have carried the same simulations but for a
homogeneous sparse network, i.e. $<k>=4$ and $\gamma \rightarrow \infty $.
As shown by the inset in Fig. 5(b), there is indeed a maximum around $\beta
_{o}\approx 1$. Before this value, the increase of gradient will
monotonically enhance the synchronizability, while after this value,
increase gradient will suppress synchronizability. As we have analyzed, too
large gradient induce the breaking effect and could suppress
synchronization. However, in the region of $\beta >\beta _{o}$, nodes belong
to the same tree are still synchronized and the network breaks into
synchronized clusters, therefore it is expected that the decrease of $A$
after $\beta _{o}$ will be slow. Since this time although the network is not
globally synchronized, strong coherence still exists within the clusters.
The simulation results of Fig. 5(b)\ also testified our analysis that the
network breaking effect is closely related to the mean degree $<k>$ and the
degree heterogeneity $\gamma $, i.e. Eq. \ref{break}. Since in Fig. 5(b),
where the substrate is densely connected SFN, we can not find the transition
phenomenon, while for the sparsely connected RN network for the inset plot,
we find the transition.

\section{ Discussion and Conclusion}

Now we are able to answer the paradox described at the beginning of this
paper: under what conditions that heterogeneous networks will be superior to
homogeneous ones in synchronizability. The key point is gradient. By
modifying the strength of the directed couplings, we can adjust the gradient
direction and weight. To make the SFN outstand, the proper way to set the
direction is making the gradient flow from the larger degree node and point
to the smaller degree node; and a convenient method to set the gradient
weight can be making it be proportional to the degree difference. If we set
the direction and weight in the opposite ways, SFN will be much difficult to
synchronize than RN. This finding thus indicates that, with proper gradient,
both two kinds of heterogeneities, the heterogeneous degree distribution and
heterogeneous weight distribution, can be used to improve network
synchronizability instead of suppression.

The advantage that SFN superior to RN becomes even clear when considering
the breaking effect. According to our finding, in achieving a stronger
synchronizability, heterogeneous network can bear larger gradient than
homogeneous network. The optimal gradient $\beta _{o}$, where the maximum
synchronizability is achieved, is closely related to the heterogeneity
exponent $\gamma $ and the mean degree $<k>$. The finding that sparse
network can not bear too much gradient may have important implications to
the function of many natural networks which have sparse links while gradient
exists, e.g., the metabolic network where $<k>\approx 7.4$ and the protein
network where $<k>\approx 2.39$ \cite{AB:2002,NEWMAN:SIAM}. Remarkably,
because most technological (e.g. the Internet and WWW) and biological
networks (e.g. the protein and neural networks) have the property of
disassortativity \cite{NEWMAN:2002}, i.e. larger degree nodes tend to
repulse from each other, the breaking effect will be more relevant in these
systems.

Another advantage enjoyed by this new scheme is that, in constructing the
coupling matrix, it only employs the local network information. More
accurately, the degree information of the node itself and its neighbors.
While schemes based on global network information, e.g. the betweenness
centrality employed in C-scheme and the oriented tree employed in Ref. \cite%
{NM:2006}, are possible for small size networks, the constructions based on
local information, e.g. the node degree employed in M-scheme and H-scheme,
could be more efficient and practical in practical. We note that in the new
scheme the setting of the gradient is determined by the information of the
neighbors, instead of the node itself. This is one of the key points making
the new scheme prominent. We also note that the proposed\ new scheme only
provide one choice in employing the gradient, other methods which may
involve complex construction rules and global network information, i.e. to
design the gradient case by case depending on the specific topology
structure, may further promote the synchronizability. But these methods may
not as convenient as this new scheme.

In conclusion, we have employed gradient network to the problem of network
synchronization. Under this framework, previous studies on asymmetric
networks can be well unified and the principles for synchronizability
enhancement become clear and systematic. Our studies not only indicates
that, comparing to homogeneous networks, scale-free networks are the natural
choice for synchronization, but also predict that, for practical networks of
low heterogeneity and small mean degree, there should exist an optimal
gradient $\beta _{o}$ where the synchronizability is maximized. We expect
this prediction to be verified by empirical findings in future.

\section{Acknowledgment}

X.G. Wang acknowledges the great hospitality of Arizona State University,
where part of the work was done during a visit. Y.-C. Lai, K. Park, and L.
Huang were supported by NSF under Grant No. ITR-0312131 and by AFOSR under
Grant No. FA9550-06-1-0024 and No. F49620-01-01-0317.

\end{document}